# Nonlinear interaction of torsional and longitudinal guided waves in hyperelastic cylinders

Yang Liu, Ehsan Khajeh, Cliff J. Lissenden [a], Joseph L. Rose

*Department of Engineering Science and Mechanics, The Pennsylvania State University, University Park,*

*Pennsylvania 16802*

(Uploaded 19 September 2012)

[a] Author to whom correspondence should be addressed. Electronic mail: cjl9@psu.edu

The nonlinear forcing terms for the wave equation in general curvilinear coordinates are derived based on a hyperelastic material. The expressions for the nonlinear part of the first Piola-Kirchhoff stress are specialized for axisymmetric torsional and longitudinal fundamental wave fields in a cylinder. The matrix characteristics of the nonlinear forcing terms and secondary mode wavestructures are manipulated to analyze the higher harmonic generation due to the guided wave mode self interactions and mutual interactions. It is proven that both torsional and longitudinal mode secondary fields can be cumulative by specific type of guided wave mode interactions. A method for the selection of preferred fundamental excitations that generate strongly cumulative higher harmonics is formulated, and described in detail for second harmonic generation. Hyperelastic finite element models are built to simulate second harmonic generation by $T(0,3)$ and $L(0,4)$ modes. A linear increase of the modal amplitude ratio $A_2 / A_1^2$ over the propagation distance is observed for both cases, which indicates mode $L(0,5)$ is effectively generated as cumulative second harmonics.

## I. INTRODUCTION

Linear ultrasonic guided waves are generally sensitive to gross defects, i.e., open cracks and large scale corrosion. However, it is highly desirable to detect damage at the smallest scale possible to maintain the best achievable structural integrity. Nonlinear ultrasonics has been shown to have the capability to provide sensitivity to micro-structural changes[1-3]. Cantrell and Yost[4] used nonlinear bulk waves to characterize fatigue damage and developed a model between the acoustic nonlinearity and dislocation dipoles. Likewise, Cantrell[5] studied the harmonic generation in cyclically stressed wavy slip metals and correlated nonlinearity parameter $\beta$ with the percentage of total life. The higher harmonic generation of guided waves in plates has been studied by Deng[6,7], de Lima and Hamilton[8], Srivastava and Lanza di Scalea[9], and Muller et al[10]. In a previous article[11], the authors made a thorough investigation of second harmonic generation in weakly nonlinear plates, for both SH and Lamb wave modes. Based on the parity analysis of the nonlinear forcing terms and secondary mode shapes, it was found that SH modes are capable of generating cumulative symmetric Lamb modes, and experimental results confirmed the theoretical predications.

Axisymmetric guided waves can propagate large distances in cylinders with little energy loss. They have been successfully applied in pipeline inspections[12-14]. However, there has been little study of higher harmonic generation of guided waves in cylindrical waveguides. De Lima and Hamilton[15] studied the higher harmonic generation of waves in waveguides having arbitrary cross-section and gave examples of cylindrical rods and shells. They solved the nonlinear wave equation by a normal mode expansion method using a perturbation analysis, and then they investigated the cumulative condition for the longitudinal modes.

Srivastava and Lanza di Scalea[16] studied the higher harmonic generation of longitudinal modes in rods, and claimed that the nature of the primary generating modes restricts the mode circumferential orders that can be generated as higher harmonics.

This work contributes by formulating the nonlinear forcing terms in general curvilinear coordinates and giving the expressions of the nonlinear part of the first Piola-Kirchhoff stress for axisymmetric torsional and longitudinal fundamental modes in particular. Higher harmonic generation due to the fundamental mode self interactions and mutual interactions are studied. The internal resonance plots for the selection of the preferred fundamental excitations are created and discussed in detail for second harmonic generation. Hyperelastic finite element models are built for axisymmetric torsional and longitudinal types of fundamental wave fields to confirm the mode selection method. Combining a cumulative harmonic with the penetration power of guided waves could be very beneficial for nondestructive evaluation and eventually structural health monitoring.

## II. NONLINEAR FORCING TERMS IN CURVILINEAR COORDINATES

Before beginning the analysis, it is worth noting that all the derivations are undertaken in the reference configuration. Assume two sets of reference coordinate systems in Euclidean space, one is Cartesian coordinates $(X^1, X^2, X^3)$ with basis vectors $(\mathbf{i}, \mathbf{j}, \mathbf{k})$, another is an arbitrary curvilinear coordinate system $(Y^1, Y^2, Y^3)$ with covariant basis vectors $(\mathbf{e}_1, \mathbf{e}_2, \mathbf{e}_3)$. The transformation between the two coordinate systems is given by $Y = X(X^1, X^2, X^3)$, and $X = Y(Y^1, Y^2, Y^3)$ is the corresponding inverse transformation.

In the absence of body forces, the equation of motion for an isotropic, homogeneous, hyperelastic waveguide[15], in a general curvilinear coordinate system, is given by

$$\mu u_{i;j}{}^{j}+(\lambda+\mu)u_{j;i}{}^{j}+\bar{f}_{i}=\rho_{0}\ddot{u}_{i}, \tag{1}$$

with the stress free boundary condition

$$\left(T_{ij}^{R-L}+\bar{T}_{ij}\right)n^{j}=0, \tag{2}$$

where $u_i = u_i\left(Y^1, Y^2, Y^3, t\right)$ is the particle displacement given in the reference configuration, $u_{i;j}$ is the covariant derivative of $u_i$ with respect to $Y^j$. $T_{ij}^{R-L}$ and $\bar{T}_{ij}$ are the linear and nonlinear parts of the first Piola-Kirchhoff stress, respectively, while $\bar{f}_i$ is the divergence of $\bar{T}_{ij}$. All of the nonlinear terms are contained in $\bar{T}_{ij}n^j$ and $\bar{f}_i$, which are often referred to as the nonlinear surface traction and the nonlinear body force, respectively. $n^j$ is the outward unit normal of the boundary surface. $\lambda$ and $\mu$ are Lamé's constants, $\rho_0$ is the referential material density.

De Lima and Hamilton[8] showed that the secondary wave field is generated due to the power flux from the fundamental wave field through nonlinear stress $\bar{T}_{ij}$ and nonlinear body force $\bar{f}_i$. Goldberg[17] provided the expressions of the nonlinear forcing terms for hyperelastic media in Cartesian coordinates. However, Cartesian tensor equations derived through partial differentiation are generally invalid in curvilinear coordinates.

The Green strain tensor in general curvilinear coordinates is given by

$$E_{ij} = \frac{1}{2}\left(u_{i;j}+u_{j;i}+u^{k}{}_{;i}u_{k;j}\right), \tag{3}$$

the covariant derivative $u_{i;j}$ is defined as

$$u_{i;j} = \frac{\partial u_i}{\partial Y^j} - u_k \Gamma_{ij}^k, \tag{4}$$

where $\Gamma_{ij}^k$ is the Euclidean Christoffel symbol, which is given by a linear combination of the derivatives of the metric tensor $g_{ij}$

$$\Gamma_{ij}^{k} = \frac{1}{2}g^{km}\left(\frac{\partial g_{mj}}{\partial Y^i}+\frac{\partial g_{im}}{\partial Y^j}-\frac{\partial g_{ij}}{\partial Y^m}\right), \tag{5}$$

the covariant metric tensor $g_{ij}$ is given by

$$g_{ij} = \delta_{mn} \frac{\partial X^m}{\partial Y^i} \frac{\partial X^n}{\partial Y^j} = \mathbf{e}_i \cdot \mathbf{e}_j, \tag{6}$$

where $\delta_{mn}$ is the Kronecker delta. $g_{ij}$ has a contravariant counterpart $g^{ij}$, which is given by its inverse.

The hyperelastic strain energy function proposed by Landau and Lifshitz [18] can be generalized to a general curvilinear coordinate system by

$$S_\mathbf{E} = \frac{1}{2}\lambda\left(E_k^{\ k}\right)^2 + \mu E_{ij} E^{ji} + \frac{1}{3}C\left(E_k^{\ k}\right)^3 + B E_k^{\ k} E_{ij} E^{ji} + \frac{1}{3} A E_{ij} E^j_{\ k} E^{ki} + O\left(\left(E_i^{\ i}\right)^4\right) \tag{7}$$

up to and including the third order terms in strain multiples where $A$, $B$ and $C$ are Landau and Lifshitz third order elastic constants. The second Piola-Kirchhoff stress $T_{ij}^{RR}$ is a referential measure of stress that is paired with the Green strain tensor through the strain energy function (7) which gives

$$\begin{aligned} T_{ij}^{RR} &= \lambda u_{l;}^{\ l} g_{ij} + \frac{\lambda}{2} u_{k;l} u^{k;l} g_{ij} + \mu\left(u_{i;j} + u_{j;i}\right) + \mu u_{k;i} u^k_{\ ;j} \\ &+ C\left(u_{l;}^{\ l}\right)^2 g_{ij} + B u_{l;}^{\ l}\left(u_{i;j} + u_{j;i}\right) + \frac{B}{2}\left(u_{l;k} u^{k;l} + u_{k;l} u^{k;l}\right) g_{ij} \\ &+ \frac{A}{4}\left(u_{i;k} u^k_{\ ;j} + u_{k;i} u_{j;}^{\ k} + u_{i;k} u_{j;}^{\ k} + u_{k;i} u^k_{\ ;j}\right) + O\left(\left(u_{i;j}\right)^3\right). \end{aligned} \tag{8}$$

The second Piola-Kirchhoff stress can be decomposed into linear part $T_{ij}^{RR-L}$ and nonlinear part $T_{ij}^{RR-NL}$, where

$$T_{ij}^{RR-L} = \lambda u_{l;}^{\ l} g_{ij} + \mu\left(u_{i;j} + u_{j;i}\right), \tag{9}$$

$$\begin{aligned} T_{ij}^{RR-NL} &= \frac{\lambda}{2} u_{k;l} u^{k;l} g_{ij} + \mu u_{k;i} u^k_{\ ;j} + C\left(u_{l;}^{\ l}\right)^2 g_{ij} \\ &+ \frac{B}{2}\left(u_{l;k} u^{k;l} + u_{k;l} u^{k;l}\right) g_{ij} + B u_{l;}^{\ l}\left(u_{i;j} + u_{j;i}\right) \\ &+ \frac{A}{4}\left(u_{i;k} u^k_{\ ;j} + u_{k;i} u_{j;}^{\ k} + u_{i;k} u_{j;}^{\ k} + u_{k;i} u^k_{\ ;j}\right) + O\left(\left(u_{i;j}\right)^3\right). \end{aligned} \tag{10}$$

The first Piola-Kirchhoff stress $T_{ij}^R$ is related to the second Piola-Kirchhoff stress $T_{ij}^{RR}$ by

$$T_{ij}^R = T_{ij}^{RR} + T_{ik}^{RR} u^k{}_{;j}. \tag{11}$$

Substituting the linear and nonlinear parts of the second Piola-Kirchhoff stress into (11), and grouping all the nonlinear terms leads to the nonlinear part of first Piola-Kirchhoff stress

$$T_{ij}^{R-NL} = T_{ij}^{RR-NL} + T_{ik}^{RR-L} u^k{}_{;j} + T_{ik}^{RR-NL} u^k{}_{;j}. \tag{12}$$

Substituting (9) and (10) into (12), the nonlinear part of first Piola-Kirchhoff stress can be obtained

$$\begin{aligned}\bar{T}_{ij} =& \frac{\lambda}{2} u_{k;l} u^{k;l} g_{ij} + (\lambda + B) u_{l;}{}^l u_{i;j} + C\left(u_{l;}{}^l\right)^2 g_{ij} \\ &+ \frac{B}{2}\left(u_{l;k} u^{k;l} + u_{k;l} u^{k;l}\right) g_{ij} + B u_{l;}{}^l u_{j;i} + \frac{A}{4} u_{k;i} u^k{}_{;j} \\ &+ \left(\mu + \frac{A}{4}\right)\left(u_{i;k} u^k{}_{;j} + u_{k;i} u^k{}_{;j} + u_{i;k} u_j{}^{;k}\right) + O\left(\left(u_{i;j}\right)^3\right).\end{aligned} \tag{13}$$

Furthermore, the nonlinear body forces are given by

$$\bar{f}_i = \bar{T}_{ij;k} g^{jk}. \tag{14}$$

Equation (13) and (14) give the nonlinear forcing terms in a general curvilinear coordinates, which can be easily adopted to a specific curvilinear coordinate system such as cylindrical coordinates by using the metric tensor and Christoffel symbols in the corresponding coordinates. The nonzero terms of the nonlinear stress in cylindrical coordinates for axisymmetric longitudinal and torsional fundamental modes are given in the Appendix.

## III CUMULATIVE HARMONIC GENERATION DUE TO MODE INTERACTIONS

Consider the interaction of two guided waves modes $\mathbf{u}^{(a)}$ and $\mathbf{u}^{(b)}$ propagating in the cylinder. The total displacement field can be decomposed into fundamental and secondary

wave fields

$$\mathbf{u} = \mathbf{u}^{(a)} + \mathbf{u}^{(b)} + \mathbf{u}^{(aa)} + \mathbf{u}^{(bb)} + \mathbf{u}^{(ab)},  \quad (15)$$

where $\mathbf{u}^{(aa)}$ and $\mathbf{u}^{(bb)}$ are the secondary wave fields due to the self interaction of mode *a* and mode *b*, respectively, and $\mathbf{u}^{(ab)}$ is the displacement field due to the mutual interaction of the two modes. Following a similar method to de Lima and Hamilton[15], the axisymmetric torsional and longitudinal mode interaction problem can be solved by the normal mode expansion technique of Auld[19]. The secondary particle velocity field $\mathbf{v}^{(2)}(R,Z,t)$ is given by

$$\mathbf{v}^{(2)}(R,Z,t) = \frac{1}{2}\sum_m \frac{i\left(p_n^{surf} + p_n^{vol}\right)}{4P_{mn}\left[k_n^* - (k_a \pm k_b)\right]} \mathbf{v}_m(R) e^{i\{(k_a \pm k_b)Z - (\omega_a \pm \omega_b)t\}} + c.c., \quad k_n^* \neq k_a \pm k_b, \quad (16)$$

$$\mathbf{v}^{(2)}(R,Z,t) = \frac{1}{2}\sum_m \frac{\left(p_n^{surf} + p_n^{vol}\right)Z}{4P_{mn}} \mathbf{v}_m(R) e^{i\{(k_a \pm k_b)Z - (\omega_a \pm \omega_b)t\}} + c.c., \quad k_n^* = k_a \pm k_b, \quad (17)$$

where

$$P_{mn} = -\frac{\pi}{8}\int_{R_i}^{R_o}\left(\mathbf{v}_n^* \cdot \mathbf{T}_m + \mathbf{v}_m \cdot \mathbf{T}_n^*\right)\cdot \mathbf{n}_Z R dR, \quad (18)$$

$$p_n^{surf} = \pi R_o \left(\mathbf{v}_n^*(R_o) \cdot \overline{\mathbf{T}}(R_o) \cdot \mathbf{n}_R\right) - \pi R_i \left(\mathbf{v}_n^*(R_i) \cdot \overline{\mathbf{T}}(R_i) \cdot \mathbf{n}_R\right), \quad (19)$$

$$p_n^{vol} = \pi \int_{R_i}^{R_o} \mathbf{v}_n^* \cdot \overline{\mathbf{f}} \cdot R dR, \quad (20)$$

where $\mathbf{v}_m$ is the particle velocity of the $m^{th}$ secondary wave mode at $\omega_a \pm \omega_b$, $P_{mn}$ is the complex power flux in the propagation direction, $k_n^*$ is the wavenumber of the unique mode such that $P_{mn} \neq 0$, $p_n^{surf}$ and $p_n^{vol}$ are interpreted as power fluxes through the surface and through the volume, respectively, due to the nonlinear surface traction and body force exerted by the fundamental waves in the cylinders. $R_i$ and $R_o$ are the inner and outer radii of the cylinder, respectively. Notice that these equations have a similar form to secondary Lamb waves in a plate[8, 11]. As indicated in equation (15), $\overline{\mathbf{T}}$ and $\overline{\mathbf{f}}$ contain terms corresponding to mode self interactions and mutual interactions. The following sections will discuss details of each case.

## A. Guided wave mode self interactions

The expressions of the nonlinear part of first Piola-Kirchhoff stress for axisymmetric longitudinal and torsional modes in the reference configuration are given in the Appendix. For both types of fundamental fields, $\bar{\mathbf{T}}$ and $\bar{\mathbf{f}}$ can be written in the matrix form as

$$\bar{\mathbf{T}} = \begin{bmatrix} \bar{T}_{RR} & 0 & \bar{T}_{RZ} \\ 0 & \bar{T}_{\Theta\Theta} & 0 \\ \bar{T}_{ZR} & 0 & \bar{T}_{ZZ} \end{bmatrix}, \tag{21}$$

$$\bar{\mathbf{f}} = \begin{Bmatrix} \bar{f}_R \\ 0 \\ \bar{f}_Z \end{Bmatrix}, \tag{22}$$

which enables the determination of the possible types of cumulative secondary wave fields due to the guided wave mode self interactions.

With regards to axisymmetric torsional mode secondary wave fields, the power flux from an arbitrary fundamental mode to a prescribed torsional secondary mode via nonlinear surface traction is given by

$$\begin{aligned} p_n^{surf-T} = \pi R_o & \left\{ \{0, v_\Theta^*(R_o), 0\} \cdot \begin{bmatrix} \bar{T}_{RR}(R_o) & 0 & \bar{T}_{RZ}(R_o) \\ 0 & \bar{T}_{\Theta\Theta}(R_o) & 0 \\ \bar{T}_{ZR}(R_o) & 0 & \bar{T}_{ZZ}(R_o) \end{bmatrix} \cdot \begin{Bmatrix} 1 \\ 0 \\ 0 \end{Bmatrix} \right\} \\ -\pi R_i & \left\{ \{0, v_\Theta^*(R_i), 0\} \cdot \begin{bmatrix} \bar{T}_{RR}(R_i) & 0 & \bar{T}_{RZ}(R_i) \\ 0 & \bar{T}_{\Theta\Theta}(R_i) & 0 \\ \bar{T}_{ZR}(R_i) & 0 & \bar{T}_{ZZ}(R_i) \end{bmatrix} \cdot \begin{Bmatrix} 1 \\ 0 \\ 0 \end{Bmatrix} \right\} = 0, \end{aligned} \tag{23}$$

additionally, the power flux resulting from the nonlinear body force is given by

$$p_n^{vol-T} = \pi \int_{R_i}^{R_o} \{0, v_\Theta^*(R), 0\} \cdot \begin{Bmatrix} \bar{f}_R(R) \\ 0 \\ \bar{f}_Z(R) \end{Bmatrix} R dR = 0, \tag{24}$$

thus the total power flux from the fundamental wave fields to a torsional type secondary wave field due to mode self interactions is

$$p_n^{total-T} = p_n^{surf-T} + p_n^{vol-T} = 0. \tag{25}$$

Considering the resulting axisymmetric longitudinal mode secondary wave field, the power flux from an arbitrary fundamental mode to a prescribed longitudinal secondary mode via the nonlinear surface traction is given by

$$\begin{aligned}
p_n^{surf-L} &= \pi R_o \left\{ \{v_R^*(R_o), 0, v_Z^*(R_o)\} \cdot \begin{bmatrix} \overline{T}_{RR}(R_o) & 0 & \overline{T}_{RZ}(R_o) \\ 0 & \overline{T}_{\Theta\Theta}(R_o) & 0 \\ \overline{T}_{ZR}(R_o) & 0 & \overline{T}_{ZZ}(R_o) \end{bmatrix} \cdot \begin{Bmatrix} 1 \\ 0 \\ 0 \end{Bmatrix} \right\} \\
&\quad - \pi R_i \left\{ \{v_R^*(R_i), 0, v_Z^*(R_i)\} \cdot \begin{bmatrix} \overline{T}_{RR}(R_i) & 0 & \overline{T}_{RZ}(R_i) \\ 0 & \overline{T}_{\Theta\Theta}(R_i) & 0 \\ \overline{T}_{ZR}(R_i) & 0 & \overline{T}_{ZZ}(R_i) \end{bmatrix} \cdot \begin{Bmatrix} 1 \\ 0 \\ 0 \end{Bmatrix} \right\} \\
&= \pi R_o \left\{ v_R^*(R_o)\overline{T}_{RR}(R_o) + v_Z^*(R_o)\overline{T}_{ZR}(R_o) \right\} \\
&\quad - \pi R_i \left\{ v_R^*(R_i)\overline{T}_{RR}(R_i) + v_Z^*(R_i)\overline{T}_{ZR}(R_i) \right\},
\end{aligned} \tag{26}$$

and the power flux resulting from the nonlinear body force is given by

$$\begin{aligned}
p_n^{vol-L} &= \pi \int_{R_i}^{R_o} \{v_R^*(R), 0, v_Z^*(R)\} \cdot \begin{Bmatrix} \overline{f}_R(R) \\ 0 \\ \overline{f}_Z(R) \end{Bmatrix} R dR \\
&= \pi \int_{R_i}^{R_o} \left( v_R^*(R)\overline{f}_R(R) + v_Z^*(R)\overline{f}_Z(R) \right) R dR.
\end{aligned} \tag{27}$$

Thus the total power flux from the fundamental wave field to the prescribed longitudinal secondary wave field due to mode self interactions is given by

$$\begin{aligned}
p_n^{total-L} &= \pi R_o \left\{ v_R^*(R_o)\overline{T}_{RR}(R_o) + v_Z^*(R_o)\overline{T}_{ZR}(R_o) \right\} \\
&\quad - \pi R_i \left\{ v_R^*(R_i)\overline{T}_{RR}(R_i) + v_Z^*(R_i)\overline{T}_{ZR}(R_i) \right\} \\
&\quad + \pi \int_{R_i}^{R_o} \left( v_R^*(R)\overline{f}_R(R) + v_Z^*(R)\overline{f}_Z(R) \right) R dR.
\end{aligned} \tag{28}$$

Equation (28) gives the power flux intensity from an arbitrary fundamental wave field to a prescribed longitudinal type secondary modes due to mode self interactions, which is in general nonzero. Thus for the case of guided wave mode self interactions: only longitudinal axisymmetric secondary wave fields can be in internal resonance with the fundamental modes. Either type of fundamental mode (longitudinal or torsional) can generate a cumulative longitudinal secondary wave field. While torsional mode secondary fields receive no power flux from either type of fundamental mode, no cumulative secondary torsional modes can be generated by a unitary excitation.

## B. Guided wave mode mutual interactions

The nonlinear forcing terms for the mode mutual interaction problems can be obtained by substituting $\mathbf{u}^{(a)}$ and $\mathbf{u}^{(b)}$ into equations (13) and (14), while retaining terms up to the second order for the nonlinear stress, and the third order for the nonlinear body forces. By doing so the nonlinear stress for the mode mutual interaction in general curvilinear coordinates is given by

$$\begin{aligned}
\bar{T}_{ij}^M &= \lambda u^{(a)\ k}_{\ \ k;} u^{(b)}_{\ \ i;j} + \lambda u^{(b)\ l}_{\ \ l;} u^{(a)}_{\ \ i;j} \\
&+ \frac{\lambda}{2}\left(u^{(a)}_{\ k;l} u^{(b)k}_{\ \ \ ;l} + u^{(b)}_{\ k;l} u^{(a)k}_{\ \ \ ;l}\right) g_{ij} \\
&+ \mu u^{(a)}_{\ i;k}\left(u^{(b)k}_{\ \ \ ;j} + u^{(b)\ k}_{\ \ j;}\right) + \mu u^{(b)}_{\ i;k}\left(u^{(a)k}_{\ \ \ ;j} + u^{(a)\ k}_{\ \ j;}\right) \\
&+ \mu\left(u^{(a)}_{\ k;i} u^{(b)k}_{\ \ \ ;j} + u^{(b)}_{\ k;i} u^{(a)k}_{\ \ \ ;j}\right) + 2C u^{(a)\ k}_{\ \ k;} u^{(b)\ l}_{\ \ l;} g_{ij} \\
&+ B u^{(a)\ k}_{\ \ k;}\left(u^{(b)}_{\ \ i;j} + u^{(b)}_{\ \ j;i}\right) + B u^{(b)\ l}_{\ \ l;}\left(u^{(a)}_{\ \ i;j} + u^{(a)}_{\ \ j;i}\right) \\
&+ \frac{B}{2}\left(u^{(a)}_{\ l;k} u^{(b)k}_{\ \ \ ;l} + u^{(b)}_{\ l;k} u^{(a)k}_{\ \ \ ;l} + u^{(a)}_{\ k;l} u^{(b)k}_{\ \ \ ;l} + u^{(b)}_{\ k;l} u^{(a)k}_{\ \ \ ;l}\right) g_{ij} \\
&+ \frac{A}{4}\Big(u^{(a)}_{\ i;k} u^{(b)k}_{\ \ \ ;j} + u^{(b)}_{\ i;k} u^{(a)k}_{\ \ \ ;j} + u^{(a)}_{\ k;i} u^{(b)\ k}_{\ \ j;} + u^{(b)}_{\ k;i} u^{(a)\ k}_{\ \ j;} \\
&\quad + u^{(a)}_{\ k;i} u^{(b)k}_{\ \ \ ;j} + u^{(b)}_{\ k;i} u^{(a)k}_{\ \ \ ;j} + u^{(a)}_{\ i;k} u^{(b)\ k}_{\ \ j;} + u^{(b)}_{\ i;k} u^{(a)\ k}_{\ \ j;}\Big) \\
&+ O\left(\left(u^{(a)}_{\ i;j}\right)^3\right) + O\left(\left(u^{(b)}_{\ i;j}\right)^3\right),
\end{aligned} \quad (29)$$

and the nonlinear body force for the mode mutual interaction problem is given by

$$\bar{f}_i^M = \bar{T}_{ij;k}^M g^{jk}. \qquad (30)$$

By checking the matrix characteristics of equations (29) and (30), two sub-cases exist for the mutual interaction between two fundamental mode excitations: (i) $\mathbf{u}^{(a)}$ and $\mathbf{u}^{(b)}$ are the same mode type, i.e., both longitudinal or both torsional modes. For this case, the matrix forms of the nonlinear forcing terms are exactly the same as equations (21) and (22). (ii) $\mathbf{u}^{(a)}$ is a different mode type from $\mathbf{u}^{(b)}$. The nonlinear forcing terms can be obtained via equations (29) and (30) by considering cylindrical coordinates with $u_R^{(a)} \neq 0$, $u_\Theta^{(a)} = 0$, $u_Z^{(a)} \neq 0$, $u_R^{(b)} = u_Z^{(b)} = 0$, $u_\Theta^{(b)} \neq 0$ (or $u_R^{(a)} = u_Z^{(a)} = 0$, $u_\Theta^{(a)} \neq 0$, $u_R^{(b)} \neq 0$, $u_\Theta^{(b)} = 0$, $u_Z^{(b)} \neq 0$) and $u_i$ is independent of $\Theta$ [20]. The matrix form of the nonlinear forcing terms for this case will be

$$\overline{\mathbf{T}}^M = \begin{bmatrix} \overline{T}_{RR}^M & \overline{T}_{R\Theta}^M & \overline{T}_{RZ}^M \\ \overline{T}_{\Theta R}^M & \overline{T}_{\Theta\Theta}^M & \overline{T}_{\Theta Z}^M \\ \overline{T}_{ZR}^M & \overline{T}_{Z\Theta}^M & \overline{T}_{ZZ}^M \end{bmatrix}, \qquad (31)$$

$$\overline{\mathbf{f}}^M = \begin{Bmatrix} \overline{f}_R^M \\ \overline{f}_\Theta^M \\ \overline{f}_Z^M \end{Bmatrix}. \qquad (32)$$

where the superscript *M* indicates the mode mutual inaction problem.

### *1.* **Longitudinal-Longitudinal (*L-L*) or Torsional-Torsional (*T-T*) mutual interactions**

When the two mutually interacting fundamental excitations are the same mode type, i.e., *L-L* or *T-T* mode mutual interactions, the nonlinear forcing terms will have the same matrix formats as mode self interaction. Thus, the cumulative second harmonic generation will also have a similar pattern as the single mode excitation, that is cumulative higher harmonics only occur for the secondary longitudinal wave field for the same mode type mutual interaction. No axisymmetric torsional secondary mode can be internally resonant given this kind of mutual interaction.

## 2. Longitudinal-Torsional (*L-T*) mutual interaction

For an axisymmetric torsional mode secondary wave field, the power flux from *L-T* mutual interaction fundamental fields to a prescribed torsional secondary mode via nonlinear surface traction is given by

$$p_n^{surf-T-M} = \pi R_o \left\{ \{0, v_\Theta^*(R_o), 0\} \cdot \begin{bmatrix} \overline{T}_{RR}^M(R_o) & \overline{T}_{R\Theta}^M(R_o) & \overline{T}_{RZ}^M(R_o) \\ \overline{T}_{\Theta R}^M(R_o) & \overline{T}_{\Theta\Theta}^M(R_o) & \overline{T}_{\Theta Z}^M(R_o) \\ \overline{T}_{ZR}^M(R_o) & \overline{T}_{Z\Theta}^M(R_o) & \overline{T}_{ZZ}^M(R_o) \end{bmatrix} \cdot \begin{Bmatrix} 1 \\ 0 \\ 0 \end{Bmatrix} \right\}$$

$$- \pi R_i \left\{ \{0, v_\Theta^*(R_i), 0\} \cdot \begin{bmatrix} \overline{T}_{RR}^M(R_i) & \overline{T}_{R\Theta}^M(R_i) & \overline{T}_{RZ}^M(R_i) \\ \overline{T}_{\Theta R}^M(R_i) & \overline{T}_{\Theta\Theta}^M(R_i) & \overline{T}_{\Theta Z}^M(R_i) \\ \overline{T}_{ZR}^M(R_i) & \overline{T}_{Z\Theta}^M(R_i) & \overline{T}_{ZZ}^M(R_i) \end{bmatrix} \cdot \begin{Bmatrix} 1 \\ 0 \\ 0 \end{Bmatrix} \right\} \quad (33)$$

$$= \pi R_o v_\Theta^*(R_o) \overline{T}_{\Theta R}^M(R_o) - \pi R_i v_\Theta^*(R_i) \overline{T}_{\Theta R}^M(R_i),$$

and the power flux via the nonlinear body forces is given by

$$p_n^{vol-T-M} = \pi \int_{R_i}^{R_o} \{0, v_\Theta^*(R), 0\} \cdot \begin{Bmatrix} \overline{f}_R^M(R) \\ \overline{f}_\Theta^M(R) \\ \overline{f}_Z^M(R) \end{Bmatrix} R dR$$

$$= \pi \int_{R_i}^{R_o} v_\Theta^*(R) \overline{f}_\Theta^M(R) R dR. \quad (34)$$

Thus the total power flux resulting from *L-T* mutual-interaction fundamental fields to a prescribed torsional secondary mode is given by

$$p_n^{total-T-M} = \pi R_o v_\Theta^*(R_o) \overline{T}_{\Theta R}^M(R_o) - \pi R_i v_\Theta^*(R_i) \overline{T}_{\Theta R}^M(R_i)$$

$$+ \pi \int_{R_i}^{R_o} v_\Theta^*(R) \overline{f}_\Theta^M(R) R dR. \quad (35)$$

The right side of equation (35) is in general nonzero, which indicates that the secondary torsional mode can be cumulative along the propagation distance provided fundamental *L-T* mutual interaction occurs, and the cumulative secondary torsional wave fields will exist at sum or difference frequencies $\omega_a \pm \omega_b$.

Considering the resulting axisymmetric longitudinal mode secondary wave field, the power flux from *L-T* mutual interaction fundamental fields to a prescribed longitudinal secondary mode via nonlinear surface traction is given by

$$p_n^{surf-L-M} = \pi R_o \left\{ \{v_R^*(R_o), 0, v_Z^*(R_o)\} \cdot \begin{bmatrix} \bar{T}_{RR}^M(R_o) & \bar{T}_{R\Theta}^M(R_o) & \bar{T}_{RZ}^M(R_o) \\ \bar{T}_{\Theta R}^M(R_o) & \bar{T}_{\Theta\Theta}^M(R_o) & \bar{T}_{\Theta Z}^M(R_o) \\ \bar{T}_{ZR}^M(R_o) & \bar{T}_{Z\Theta}^M(R_o) & \bar{T}_{ZZ}^M(R_o) \end{bmatrix} \cdot \begin{Bmatrix} 1 \\ 0 \\ 0 \end{Bmatrix} \right\}$$

$$- \pi R_i \left\{ \{v_R^*(R_i), 0, v_Z^*(R_i)\} \cdot \begin{bmatrix} \bar{T}_{RR}^M(R_i) & \bar{T}_{R\Theta}^M(R_i) & \bar{T}_{RZ}^M(R_i) \\ \bar{T}_{\Theta R}^M(R_i) & \bar{T}_{\Theta\Theta}^M(R_i) & \bar{T}_{\Theta Z}^M(R_i) \\ \bar{T}_{ZR}^M(R_i) & \bar{T}_{Z\Theta}^M(R_i) & \bar{T}_{ZZ}^M(R_i) \end{bmatrix} \cdot \begin{Bmatrix} 1 \\ 0 \\ 0 \end{Bmatrix} \right\} \quad (36)$$

$$= \pi R_o \left\{ v_R^*(R_o) \bar{T}_{RR}^M(R_o) + v_Z^*(R_o) \bar{T}_{ZR}^M(R_o) \right\}$$

$$- \pi R_i \left\{ v_R^*(R_i) \bar{T}_{RR}^M(R_i) + v_Z^*(R_i) \bar{T}_{ZR}^M(R_i) \right\},$$

and the power flux via the nonlinear body forces is given by

$$p_n^{vol-L-M} = \pi \int_{R_i}^{R_o} \{v_R^*(R), 0, v_Z^*(R)\} \cdot \begin{Bmatrix} \bar{f}_R^M(R) \\ \bar{f}_\Theta^M(R) \\ \bar{f}_Z^M(R) \end{Bmatrix} RdR$$

$$= \pi \int_{R_i}^{R_o} \left( v_R^*(R) \bar{f}_R^M(R) + v_Z^*(R) \bar{f}_Z^M(R) \right) RdR. \quad (37)$$

Thus the total power flux resulting from an arbitrary *L-T* mutual interaction fundamental field to a prescribed longitudinal secondary mode is given by

$$p_n^{total-L-M} = \pi R_o \left\{ v_R^*(R_o) \bar{T}_{RR}^M(R_o) + v_Z^*(R_o) \bar{T}_{ZR}^M(R_o) \right\}$$

$$- \pi R_i \left\{ v_R^*(R_i) \bar{T}_{RR}^M(R_i) + v_Z^*(R_i) \bar{T}_{ZR}^M(R_i) \right\} \quad (38)$$

$$+ \pi \int_{R_i}^{R_o} \left( v_R^*(R) \bar{f}_R^M(R) + v_Z^*(R) \bar{f}_Z^M(R) \right) RdR.$$

In summary, cumulative higher harmonics can be generated by either mode self interactions or mode mutual interactions. Only longitudinal secondary wave fields can be cumulative for mode self interactions, *T-T* mutual interaction or *L-L* mutual interaction. However, both torsional and longitudinal secondary wave fields can be cumulative for *L-T* mutual interaction. Table I lists all the possible cumulative secondary fields due to mode interactions.

**IV EXCITATION OF STRONG CUMULATIVE SECOND HARMONICS**

In this section, a method is illustrated for the selection of fundamental excitations that generate strong cumulative second harmonics. This method can be generalized for the selection of other order of strongly cumulative higher harmonic generations. The generation of second harmonics is due to mode self interactions. The amplitude of the second harmonic grows linearly with propagation distance, as indicated by Eqn. (17), provided it has twice the wave number of the fundamental mode (thus the phase velocities are synchronized) and $p_n^{surf} + p_n^{vol} \neq 0$ (thus the power flux is nonzero). This is known as the internal resonance condition for the second harmonics, and this cumulative nature of second harmonics has a practical reception advantage. Hence, fundamental wave modes that generate strongly cumulative second harmonics with reasonably good excitability will be sought. The dimensions and properties of a steel cylinder analyzed here for illustrative purposes are given in Table II.

Figure 1 (a) and (b) have the marked synchronism points for fundamental torsional and longitudinal wave fields, respectively. Dispersion curves for potential cumulative second harmonics are plotted with black symbols, while the dispersion curves for the fundamental wave fields are color scaled. The synchronism points are the intersection points of the fundamental and secondary wave fields. A limited number of synchronism points occur when the phase velocities are equal to: (i) the longitudinal wave speed $C_L$, (ii) the shear wave speed $C_T$, (iii) the Rayleigh wave speed $C_R$ (iv) the Lamé mode wave speed $C_{Lamé}(\sqrt{2}C_T)$, or (v) the crossing points of fundamental longitudinal modes. Matsuda and Biwa[21] and Liu et al.[11] manipulated the dispersion relations for plates to show mathematically that the synchronism points occur at any of the aforementioned conditions. While that type of analysis is not performed here, Figures 1 (a) and (b) graphically demonstrate that the same synchronism points exist for axisymmetric guided modes in cylinders.

Figures 1 (a) and (b) are termed as internal resonance plots, because they exhibit both the synchronism points and the normalized power flux intensity (obtained by Eqn. 28) from the fundamental modes to a prescribed secondary wave field, $L(0,5)$ in this case. The internal resonance plots enable the selection of synchronism points that have higher power flux to a prescribed secondary wave field, which implies the preferred fundamental excitations. Table III lists the power flux from fundamental excitation points to corresponding secondary wave fields for some synchronism mode pairs. Notice that high power flux occurs from the longitudinal fundamental modes to $L(0,5)$ and $L(0,9)$, while low power flux occurs to $L(0,3)$, $L(0,6)$, and $L(0,8)$. $L(0,5)$ and $L(0,9)$ correspond to symmetric Lamb modes in plate, while $L(0,3)$, $L(0,6)$, and $L(0,8)$ correspond to antisymmetric Lamb modes. In plates, antisymmetric secondary modes receive no power flux due to their symmetry properties[10, 11]. However, in cylinders the mode shapes are neither exactly symmetric or antisymmetric, thus nonzero power flux occurs to all longitudinal modes, albeit it is small to those modes that correspond to antisymmetric Lamb modes in plate. This analysis holds for torsional fundamental modes as well. The synchronism points with higher power flux provide stronger potential for nonlinear ultrasonic guided wave NDE of cylindrical structures. Due to this consideration of the power flux and the modal excitability, $T(0,3)$-$L(0,5)$ and $L(0,4)$-$L(0,5)$ mode pairs are selected for finite element simulation.

## V. FINITE ELEMENT SIMULATION OF HYPERELASTIC CYLINDERS

In order to demonstrate that internal resonance plot enable selection of preferred fundamental excitations, finite element simulations have been performed. Hyperelastic finite element models with constitutive equations (3) and (7) to (14) have been constructed. In order to guarantee convergent solutions in the time domain and adequately accurate solutions in the spatial domain during wave propagation simulations of second harmonic generation, two crucial criteria must be satisfied: (i) the maximum length of each element in the wave propagation direction should be less than $\lambda_{2f}/8$, where $\lambda_{2f}$ is the wavelength of the secondary mode, (ii) the time step $\Delta t$ must be selected according to the ratio of the smallest element length $\Delta L_{\min}$ in the wave propagation direction and the fastest group velocity $C_g$ that exists in the simulation, i.e., $\Delta t \leq \Delta L_{\min}/C_g$.

### A. Axisymmetric longitudinal mode fundamental excitation

The simulations enable identification of the second harmonic generation and its cumulative tendency. A steel pipe with the parameters given in Table II is modeled. A 4-element inter digital transducer (IDT) that wraps around the circumference of the pipe has been simulated by applying surface tractions. A 15-cycle tone burst excitation of the axial traction component with a central frequency of $f_0 = 2.569$ *MHz* was applied in order to excite the *L*(0,4) mode at the longitudinal wave speed. The IDT element spacing matches the half wavelength and the element width is half of the spacing. For the *L*(0,4) mode at $f_0 = 2.569$ *MHz* the wavelength is 2.318 *mm*.

All of the measurements have been made in the far field, at propagation distances from 45 *mm* to 95 *mm* with an increment of 12.5 *mm*. The time domain signals received for a propagation distance of 57.5 *mm* is shown in Fig. 2(a). A Tukey window has been applied to extract the first received signal for the Fourier transform. Figure 2(b) shows the frequency spectrum of the fundamental *L*(0,4) excitation. The modal amplitude ratio, $A_2 / A_1^2$, where $A_1$ and $A_2$ are the amplitudes of the fundamental *L*(0,4) mode (at $f_0$) and secondary *L*(0,5) mode (at $2f_0$), respectively, is plotted as a function of the propagation distance in Fig. 3. It is evident from Fig. 3 that the modal amplitude ratio $A_2 / A_1^2$ (which is a relative measure of the nonlinearity parameter $\beta$) for *L*(0,4) fundamental excitation increases linearly with propagation distance, thus the *L*(0,4)-*L*(0,5) mode pair is internally resonant.

**B. Axisymmetric torsional mode fundamental excitation**

A second finite element model applies the fundamental *T*(0,3) mode excitation to the same steel pipe. A 4-element IDT has been used for this simulation. The elements in the IDT are spaced at the half wavelength (1.159 *mm*), and their widths are half of the spacing. A tone burst traction with a central frequency 2.569 *MHz* has been applied in the circumferential direction for the excitation of the fundamental *T*(0,3) wave field.

Figure 4 shows the received time domain signals from the *T*(0,3) fundamental excitation for a 57.5 *mm* propagation distance. The black lines correspond to the circumferential component of the displacement field $u_\Theta$ in the simulation, which represents the displacement of the fundamental *T*(0,3) mode. The red lines correspond to the axial component of the displacement field $u_Z$, which is the *L*(0,5) mode due to the second harmonic generation. Since the *T*(0,3) and *L*(0,5) wave packages arrive at nearly the same time, this indicates that their group velocities are very similar, and can be deemed to be matched.

Figure 5 shows the frequency spectrum of the $u_\Theta$ and $u_Z$ received signals from the $T(0,3)$ fundamental excitation. As in Fig. 4 the axis scales for $u_\Theta$ (shown in black) and $u_Z$ (shown in red) are significantly different. The $u_\Theta$ signal appears at the fundamental (2.569 *MHz*) harmonic frequency, while the $u_Z$ signal is present at the second and fourth harmonic frequencies. As indicated by Eqn. (25), no $T(0,3)$ second harmonic is generated; while as indicated by Eqn. (28) a $L(0,5)$ second harmonic is generated.

In the same fashion that $A_2/A_1^2$ is plotted as a function of propagation length for longitudinal modes in Fig. 3, the modal amplitude ratio, $A_2/A_1^2$, where $A_1$ and $A_2$ are the amplitudes of the fundamental $T(0,3)$ mode ($u_\Theta$ at $f_0$) and secondary $L(0,5)$ mode ($u_Z$ at $2f_0$) respectively, is plotted as a function of the propagation distance in Fig.6. The $T(0,3)$ to $L(0,5)$ mode pair is internally resonant and $A_2/A_1^2$ increases linearly with propagation distances as expected.

## VI. CONCLUSION

The generation of higher harmonics due to the mode interactions of axisymmetric torsional and longitudinal modes in isotropic, homogenous, hyperelastic cylinders has been investigated. Strongly cumulative higher harmonic generation is achieved by satisfying (i) high power flux from fundamental mode to the secondary wave field and (ii) the synchronism condition. For higher harmonic generation of axisymmetric modes in cylindrical waveguides we conclude:

**A.** Cumulative axisymmetric torsional mode secondary wave fields can only be generated by fundamental longitudinal-torsional mutual interaction, and it will present at sum and difference frequencies. Secondary longitudinal modes can be cumulative over the propagation length for any type of fundamental wave field. Both axisymmetric longitudinal and torsional mode fundamental fields have the potential to generate cumulative secondary wave fields. The analysis of axisymmetric longitudinal modes agrees with that of de Lima and Hamilton[15], as well as Srivastava and Lanza di Scalea[16], while the analysis of torsional mode self interaction and torsional-longitudinal mode mutual interactions are new.

**B.** Synchronism points for second harmonic generation have been identified graphically by plotting both fundamental and secondary mode dispersion curves. Only a finite number of fundamental excitation points meet the synchronism condition, all of which are points with special phase velocities (longitudinal wave speed $C_L$, shear wave speed $C_T$, Rayleigh wave speed $C_R$ or Lamé wave speed $C_{Lamé}$), or the crossing points of fundamental wave modes.

**C.** Internal resonance plots that explicitly present synchronism points and power flux intensity to a prescribed secondary wave field have been created. These plots aid mode and frequency selection of fundamental modes capable of generating strongly cumulative second harmonics. Preferred fundamental excitation can be achieved by selecting the synchronism points in high power flux intensity areas.

**D.** Hyperelastic finite element analysis has demonstrated that the $L(0,4)$-$L(0,5)$ mode pair at the longitudinal wave speed, which is internally resonant, results in a cumulative second harmonic generation that increases linearly over propagating distances of 45-95 *mm*.

**E.** The $T(0,3)$ mode at the longitudinal wave speed has been excited in a finite element simulation. The normalized amplitude of the $L(0,5)$ mode at the double frequency increases along the propagation distance, which means that the $T(0,3)$-$L(0,5)$ mode pair at the longitudinal wave speed is an internal resonance point. This is the first time a torsional mode fundamental excitation has been used to generate a cumulative second harmonic in a cylindrical waveguide.

**ACKNOWLEDGEMENTS**

This research is being performed using funding received from the DOE Office of Nuclear Energy's Nuclear Energy University Program under Award No. 120237.

**APPENDIX: NONLINEAR FORCING TERMS IN CYLINDRICAL COORDINATES**

In cylindrical coordinates, that is $(Y^1, Y^2, Y^3) = (R, \Theta, Z)$, the nonzero components of the covariant metric tensor and the Christoffel symbols are

$$g_{11} = 1, \ g_{22} = R^2, \ g_{33} = 1 \tag{A1}$$

$$\Gamma^1_{22} = -R, \ \Gamma^2_{12} = \Gamma^2_{21} = \frac{1}{R}. \tag{A2}$$

The nonlinear forcing terms for axisymmetric longitudinal modes can be obtained via equations (13) and (14) by considering the fundamental displacement field $u_i$ to have zero circumferential component and to be independent of $\Theta$ [20]. The nonzero terms of the nonlinear part of first Piola-Kirchhoff stress for axisymmetric longitudinal mode fundamental wave fields are given by

$$\begin{aligned}
\bar{T}_{RR} &= \left(\frac{3\lambda}{2} + 3\mu + A + 3B + C\right)\frac{\partial u_R}{\partial R}\frac{\partial u_R}{\partial R} + \left(\frac{\lambda}{2} + B + C\right)\frac{\partial u_Z}{\partial Z}\frac{\partial u_Z}{\partial Z} \\
&+ \left(\frac{\lambda}{2} + B + C\right)\left(\frac{u_R}{R}\frac{u_R}{R} + 2\frac{\partial u_R}{\partial R}\frac{u_R}{R} + 2\frac{\partial u_R}{\partial R}\frac{\partial u_Z}{\partial Z}\right) + 2C\frac{u_R}{R}\frac{\partial u_Z}{\partial Z} \\
&+ \left(\frac{\lambda}{2} + \mu + \frac{A}{4} + \frac{B}{2}\right)\left(\frac{\partial u_R}{\partial Z}\frac{\partial u_R}{\partial Z} + \frac{\partial u_Z}{\partial R}\frac{\partial u_Z}{\partial R}\right) + \left(\mu + \frac{A}{2} + B\right)\frac{\partial u_R}{\partial Z}\frac{\partial u_Z}{\partial R},
\end{aligned} \tag{A3}$$

$$\begin{aligned}
\bar{T}_{RZ} &= \left(\mu + \frac{A}{2} + B\right)\left(\frac{\partial u_R}{\partial R}\frac{\partial u_R}{\partial Z} + \frac{\partial u_R}{\partial Z}\frac{\partial u_Z}{\partial Z}\right) + (\lambda + B)\frac{\partial u_Z}{\partial R}\frac{u_R}{R} + B\frac{\partial u_R}{\partial Z}\frac{u_R}{R} \\
&+ \left(\lambda + 2\mu + \frac{A}{2} + B\right)\left(\frac{\partial u_R}{\partial R}\frac{\partial u_Z}{\partial R} + \frac{\partial u_Z}{\partial R}\frac{\partial u_Z}{\partial Z}\right),
\end{aligned} \tag{A4}$$

$$\bar{T}_{\Theta\Theta} = \left(\frac{3\lambda}{2} + 3\mu + A + 3B + C\right)\frac{u_R}{R}\frac{u_R}{R} + B\frac{\partial u_R}{\partial Z}\frac{\partial u_Z}{\partial R}$$
$$+ \left(\frac{\lambda}{2} + B + C\right)\left(\frac{\partial u_R}{\partial R}\frac{\partial u_R}{\partial R} + 2\frac{\partial u_R}{\partial R}\frac{u_R}{R} + 2\frac{\partial u_Z}{\partial Z}\frac{u_R}{R} + \frac{\partial u_Z}{\partial Z}\frac{\partial u_Z}{\partial Z}\right) \quad \text{(A5)}$$
$$+ \left(\frac{\lambda}{2} + \frac{B}{2}\right)\left(\frac{\partial u_R}{\partial Z}\frac{\partial u_R}{\partial Z} + \frac{\partial u_Z}{\partial R}\frac{\partial u_Z}{\partial R}\right) + 2C\frac{\partial u_R}{\partial R}\frac{\partial u_Z}{\partial Z},$$

$$\bar{T}_{ZR} = \left(\mu + \frac{A}{2} + B\right)\left(\frac{\partial u_R}{\partial R}\frac{\partial u_Z}{\partial R} + \frac{\partial u_Z}{\partial R}\frac{\partial u_Z}{\partial Z}\right) + (\lambda + B)\frac{\partial u_R}{\partial Z}\frac{u_R}{R}$$
$$+ \left(\lambda + 2\mu + \frac{A}{2} + B\right)\left(\frac{\partial u_R}{\partial Z}\frac{\partial u_R}{\partial R} + \frac{\partial u_R}{\partial Z}\frac{\partial u_Z}{\partial Z}\right) + B\frac{\partial u_Z}{\partial R}\frac{u_R}{R}, \quad \text{(A6)}$$

$$\bar{T}_{ZZ} = \left(\frac{3\lambda}{2} + 3\mu + A + 3B + C\right)\frac{\partial u_Z}{\partial Z}\frac{\partial u_Z}{\partial Z} + \left(\mu + \frac{A}{2} + B\right)\frac{\partial u_R}{\partial Z}\frac{\partial u_Z}{\partial R}$$
$$+ \left(\frac{\lambda}{2} + \mu + \frac{A}{4} + \frac{B}{2}\right)\left(\frac{\partial u_R}{\partial Z}\frac{\partial u_R}{\partial Z} + \frac{\partial u_Z}{\partial R}\frac{\partial u_Z}{\partial R}\right) + 2C\frac{\partial u_R}{\partial R}\frac{u_R}{R} \quad \text{(A7)}$$
$$+ \left(\frac{\lambda}{2} + B + C\right)\left(\frac{\partial u_R}{\partial R}\frac{\partial u_R}{\partial R} + \frac{u_R}{R}\frac{u_R}{R} + 2\frac{\partial u_R}{\partial R}\frac{\partial u_Z}{\partial Z} + 2\frac{u_R}{R}\frac{\partial u_Z}{\partial Z}\right).$$

The nonlinear forcing terms for axisymmetric torsional mode fundamental wave fields are obtained through equations (13) and (14) by letting $u_R = u_Z = 0$, $u_\Theta \neq 0$, and requiring $u_i$ to be independent of $\Theta$ [20]. The nonzero terms of the nonlinear stress for axisymmetric torsional mode fundamental wave fields are given by

$$\bar{T}_{RR} = \left(\frac{\lambda}{2} + \mu + \frac{A}{4} + \frac{B}{2}\right)\left(\frac{\partial u_\Theta}{\partial R}\frac{\partial u_\Theta}{\partial R} + \frac{u_\Theta}{R}\frac{u_\Theta}{R}\right)$$
$$+ \left(\frac{\lambda}{2} + \frac{B}{2}\right)\frac{\partial u_\Theta}{\partial Z}\frac{\partial u_\Theta}{\partial Z} - \left(\mu + \frac{A}{2} + B\right)\frac{\partial u_\Theta}{\partial R}\frac{u_\Theta}{R}, \quad \text{(A8)}$$

$$\bar{T}_{RZ} = \left(\mu + \frac{A}{4}\right)\frac{\partial u_\Theta}{\partial R}\frac{\partial u_\Theta}{\partial Z} - \frac{A}{4}\frac{\partial u_\Theta}{\partial Z}\frac{u_\Theta}{R}, \quad \text{(A9)}$$

$$\bar{T}_{\Theta\Theta} = \left(\frac{\lambda}{2} + \mu + \frac{A}{4} + \frac{B}{2}\right)\left(\frac{\partial u_\Theta}{\partial R}\frac{\partial u_\Theta}{\partial R} + \frac{u_\Theta}{R}\frac{u_\Theta}{R} + \frac{\partial u_\Theta}{\partial Z}\frac{\partial u_\Theta}{\partial Z}\right) - \left(\mu + \frac{A}{2} + B\right)\frac{\partial u_\Theta}{\partial R}\frac{u_\Theta}{R}, \quad \text{(A10)}$$

$$\bar{T}_{ZR} = \left(\mu + \frac{A}{4}\right)\frac{\partial u_\Theta}{\partial R}\frac{\partial u_\Theta}{\partial Z} - \left(\mu + \frac{A}{4}\right)\frac{\partial u_\Theta}{\partial Z}\frac{u_\Theta}{R}, \quad \text{(A11)}$$

$$\bar{T}_{ZZ} = \left(\frac{\lambda}{2} + \frac{B}{2}\right)\left(\frac{\partial u_\Theta}{\partial R}\frac{\partial u_\Theta}{\partial R} + \frac{u_\Theta}{R}\frac{u_\Theta}{R}\right) + \left(\frac{\lambda}{2} + \mu + \frac{A}{4} + \frac{B}{2}\right)\frac{\partial u_\Theta}{\partial Z}\frac{\partial u_\Theta}{\partial Z} - B\frac{\partial u_\Theta}{\partial R}\frac{u_\Theta}{R}. \quad \text{(A12)}$$

**REFERENCES**

[1] G. E. Dace, R. B. Thompson, L. J. H. Brasche, D. K. Rehbein, O. Buck, " Nonlinear acoustics, a technique to detection microstructural changes in materials," Rev. Prog. Quant. Nondestr. Eval. **10B**, 1685-1692 (1991).

[2] J. H. Cantrell, "Nonlinear dislocation dynamics at ultrasonic frequencies," J. Appl. Phys. **106**, 043520-7 (2009).

[3] J. Y. Kim, L. J. Jacobs, J. Qu, and J. W. Littles, "Experimental characterization of fatigue damage in a nickel-base superalloy using nonlinear ultrasonic waves," J. Acoust. Soc. Am. **120**, 1266-1273 (2006).

[4] J. H. Cantrell and W. T. Yost, "Nonlinear ultrasonic characterization of fatigue microstructures," Int. J. Fatigue **23**, 487-490 (2001).

[5] J. H. Cantrell, "Substructural organization, dislocation plasticity and harmonic generation in cyclically stressed wavy slip metals," Proc. R. Soc. London, Ser. A **460**, 757-780 (2004).

[6] Deng, M, "Cumulative second-harmonic generation accompanying nonlinear shear horizontal mode propagation in a solid plate," J. Appl. Phys. **84**: 3500-3505 (1998).

[7] Deng, M, "Cumulative second-harmonic generation of Lamb-mode propagation in a solid plate," J. Appl. Phys. **85**: 3051-3058 (1999).

[8] W. J. N. de Lima and M. F. Hamilton, "Finite-amplitude waves in isotropic elastic plates," J. Sound. Vib. **265**: 819-839 (2003).

[9] A. Srivastava and F. Lanza di Scalea, "On the existence of antisymmetric or symmetric Lamb waves at non-linear higher harmonics," J. Sound. Vib. **323**: 932-943 (2009).

[10] M. F. Muller, J. Y. Kim, J. Qu, L. J. Jacobs, "Characteristics of second harmonic

generation of Lamb waves in nonlinear elastic plates," J. Acoust. Soc. Am. **127**: 2141-2152 (2010).

[11] Y. Liu, V. K. Chillara, C. J. Lissenden, "On selection of primary modes for generation of strong internally resonant second harmonics in plates." Submitted (2012).

[12] M. J. S. Lowe, D. N. Alleyne, P. Cawley, "Defect detection in pipes using guided waves." Ultrasonics, **36**: 147-154 (1998).

[13] H. Kwun, K. A. Bartels, "Magnetostrictive sensor technology and its applications." Ultrasonics, **36**: 171-178 (1998).

[14] J. Li, and J. L. Rose. "Implementing Guided Wave Mode Control by Use of a Phased Transducer Array." IEEE Trans. Ultrason., Ferroelect., Freq. Contr. **49**: 1720-1729 (2001).

[15] W. J. N. de Lima and M. F. Hamilton, "Finite amplitude waves in isotropic elastic waveguides with arbitrary constant cross-sectional area," Wave Motion. **41**: 1-11 (2005).

[16] A. Srivastava and F. Lanza di Scalea, "On the existence of longitudinal or flexural waves in rods at nonlinear higher harmonics," J. Sound. Vib. **329**: 1499-1506 (2010).

[17] Z. Goldberg, "Interaction of plane longitudinal and transverse elastic waves," Sov. Phys. Acoust. **6**, 306–310 (1961).

[18] L. D. Landau and E. M. Lifshitz, *Theory of elasticity* (Pergamon, New York, 1986).

[19] B. A. Auld. *Acoustic fields and waves in solids* (Robert E. Krieger Publishing Company, Malabar, Florida, 1990).

[20] J. L. Rose. *Ultrasonic waves in solid media* (Cambridge University Press, Cambridge, 1999).

[21] N. Matsuda and S. Biwa, "Phase and group velocity matching for cumulative harmonic

generation in Lamb waves," J. Appl. Phys. **109**, 094903-11 (2011).

[22] J. J. Rushchitsky, "Quadratically nonlinear cylindrical hyperelastic waves: primary analysis of evolution," Int. Appl. Mech. **41**: 770-777 (2005).

[23] A. N. Norris, Finite amplitude waves in solids, in: M. F. Hamilton, D. T. Blackstock (EDs.), *Nonlinear Acoustics* (Academic Press, New York, 1998).

**LIST OF ILLUSTRATIONS**

Table I. Possible cumulative secondary wave fields due to axisymmetric mode interactions in cylinders.

Table II. Input parameters for the finite element models.

Table III. Normalized power flux intensity of some synchronism mode pairs for a steel pipe.

FIG. 1. (Color online) Internal resonance plots: (a) fundamental axisymmetric torsional modes to secondary $L(0,5)$, (b) fundamental axisymmetric longitudinal modes to secondary $L(0,5)$. Color scaled lines represent the fundamental modes superimposed with normalized power flux intensity, the black lines are the possible cumulative secondary modes. The small circles indicate the synchronism points.

FIG. 2. Signals from $L(0,4)$ fundamental excitation: (a) Received time domain signals at propagation distance of 57.5 *mm*. (b) Frequency spectrum for the first receiving wave packet, where $A_1$ and $A_2$ are fundamental and second harmonic generations, respectively.

FIG. 3. The modal amplitude ratio $A_2 / A_1^2$ for $L(0,4)$ mode excitation as a function of propagation distance. $A_1$ is the modal amplitude of $L(0,4)$ at fundamental frequency, while $A_2$ is the secondary $L(0,5)$ amplitude at double frequency. The dotted line represents the linear regression of the data points.

FIG. 4. (Color online) Time domain signals from $T(0,3)$ fundamental excitation at propagation distance of 57.5 *mm*. The black line is the displacement $u_\Theta$ component and the red line is the $u_z$ component.

FIG. 5. (Color online) Frequency spectrums from $T(0,3)$ fundamental excitation: the black line shows the FFT result of the first receiving wave packet of the $u_\Theta$ displacement

component, while the red dotted line indicates the frequency spectrum of the $u_z$ component.

FIG. 6. The modal amplitude ratio $A_2/A_1^2$ for $T(0,3)$ mode excitation as a function of propagation distance. $A_1$ is the modal amplitude of $T(0,3)$ at fundamental frequency, while $A_2$ is the secondary $L(0,5)$ amplitude at double frequency. A best fit line of the data points is included.

Table I. Possible cumulative secondary wave fields due to axisymmetric mode interactions in cylinders.

| FUNDAMENTAL WAVE FIELD | | SECONDARY WAVE MODE | |
| --- | --- | --- | --- |
| | | TORSIONAL | LONGITUDINAL |
| SELF INTERACTION | $L$ | N | Y |
| | $T$ | N | Y |
| MUTUAL INTERACTION | $T$-$T$ | N | Y |
| | $L$-$L$ | N | Y |
| | $L$-$T$ | Y | Y |

a. '$L$' represents for longitudinal mode and '$T$' means torsional mode.

b. 'Y' means the secondary mode is cumulative, 'N' indicates that the secondary field is not cumulative.

Table II. Input parameters for the finite element models.

| Dimensions | | Material properties[a] | | | | | |
|---|---|---|---|---|---|---|---|
| $R_i$ (mm) | $R_o$ (mm) | $\rho$ (kg/m$^3$) | $\lambda$ (GPa) | $\mu$ (GPa) | $A$ (GPa) | $B$ (GPa) | $C$ (GPa) |
| 9 | 10.5 | 7932 | 116.2 | 82.7 | -325 | -310 | -800 |

a. *A, B* and *C* are from the Ref. 22. A conversion between Landau and Lifshitz third order elastic constants and Murnaghan constants can be found in Morris[23].

Table III. Normalized power flux intensity of some synchronism mode pairs for a steel pipe

| Synchronism fundamental mode types | | Secondary wave field | $fd$ value ($MHz\text{-}mm$) | Normalized Power flux intensity |
|---|---|---|---|---|
| Torsional modes | $C_T$ — $T(0,1)$ | $L(0,2)$ | 1.72 | $1.29\times10^2$ |
| | $C_{Lamé}$ — $T(0,2)$ | $L(0,3)$ | 2.28 | $1.23\times10^1$ |
| | $C_{Lamé}$ — $T(0,3)$ | $L(0,6)$ | 4.56 | $2.36\times10^2$ |
| | $C_L$ — $T(0,2)$ | $L(0,4)$ | 1.92 | $4.10\times10^3$ |
| | $C_L$ — $T(0,3)$ | $L(0,5)$ | 3.85 | $1.27\times10^4$ |
| | $C_L$ — $T(0,4)$ | $L(0,7)$ | 5.78 | $5.47\times10^4$ |
| Longitudinal modes | $C_{Lamé}$ — $L(0,2)$ | $L(0,3)$ | 2.28 | $1.35\times10^{-3}$ |
| | $C_{Lamé}$ — $L(0,3)$ | $L(0,6)$ | 4.56 | $2.18\times10^{-2}$ |
| | $C_{Lamé}$ — $L(0,4)$ | $L(0,8)$ | 6.84 | $3.63\times10^2$ |
| | $C_L$ — $L(0,4)$ | $L(0,5)$ | 3.85 | $2.81\times10^4$ |
| | $C_L$ — $L(0,5)$ | $L(0,9)$ | 7.70 | $1.36\times10^5$ |

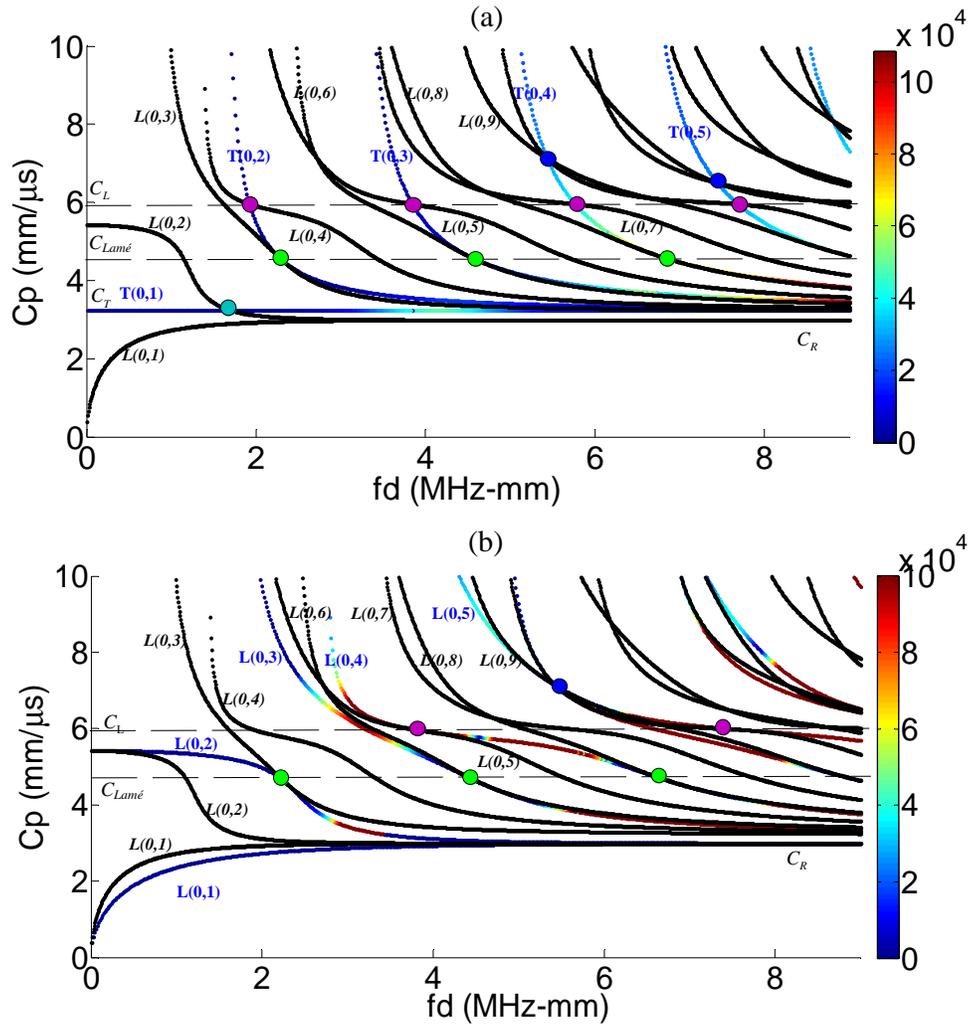

FIG. 1. (Color online) Internal resonance plots: (a) fundamental axisymmetric torsional modes to secondary $L(0,5)$, (b) fundamental axisymmetric longitudinal modes to secondary $L(0,5)$. Color scaled lines represent the fundamental modes superimposed with normalized power flux intensity, the black lines are the possible cumulative secondary modes. The small circles indicate the synchronism points.

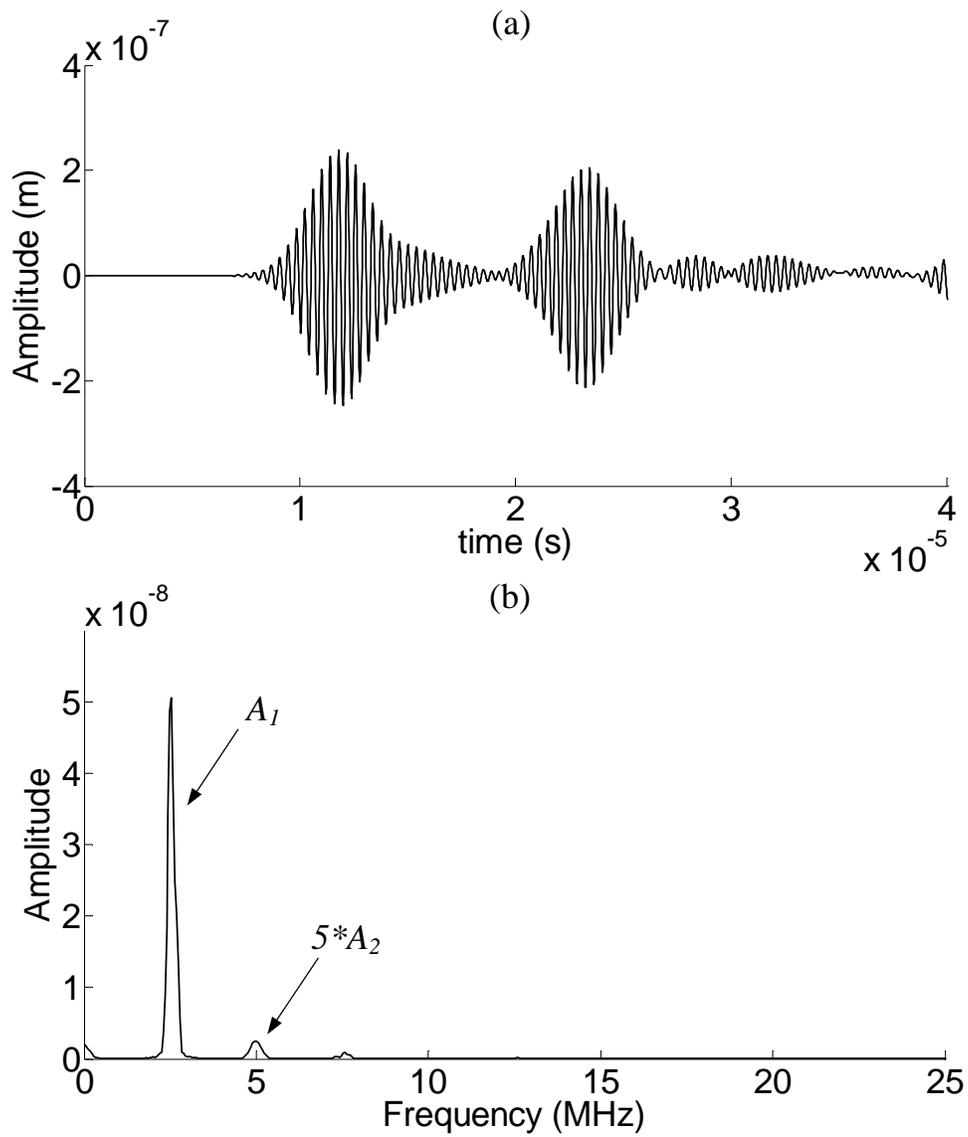

FIG. 2. Signals from $L(0,4)$ fundamental excitation: (a) Received time domain signals at propagation distance of 57.5 *mm*. (b) Frequency spectrum for the first receiving wave packet, where $A_1$ and $A_2$ are fundamental and second harmonic generations, respectively.

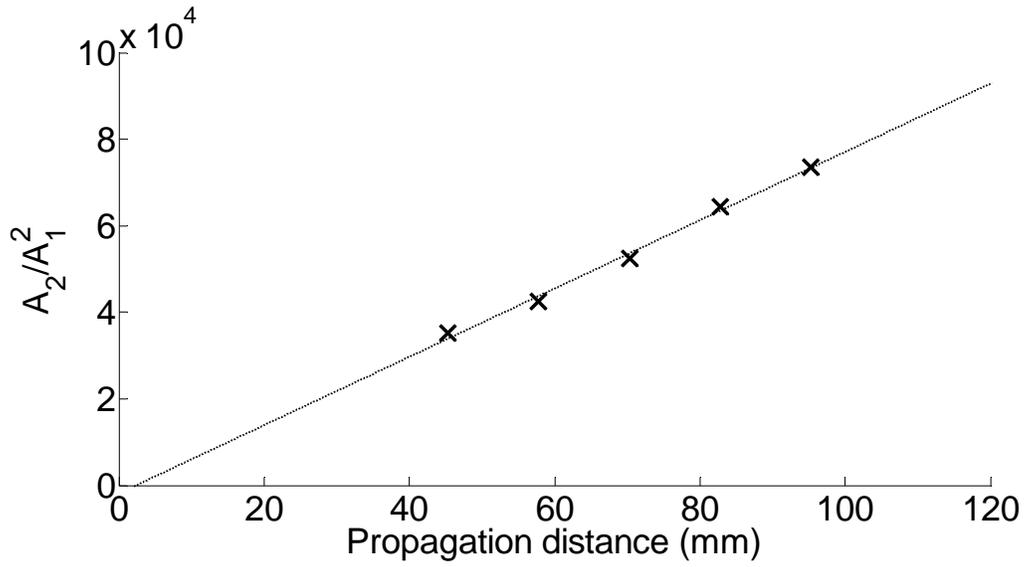

FIG. 3. The modal amplitude ratio $A_2 / A_1^2$ for $L(0,4)$ mode excitation as a function of propagation distance. $A_1$ is the modal amplitude of $L(0,4)$ at fundamental frequency, while $A_2$ is the secondary $L(0,5)$ amplitude at double frequency. The dotted line represents the linear regression of the data points.

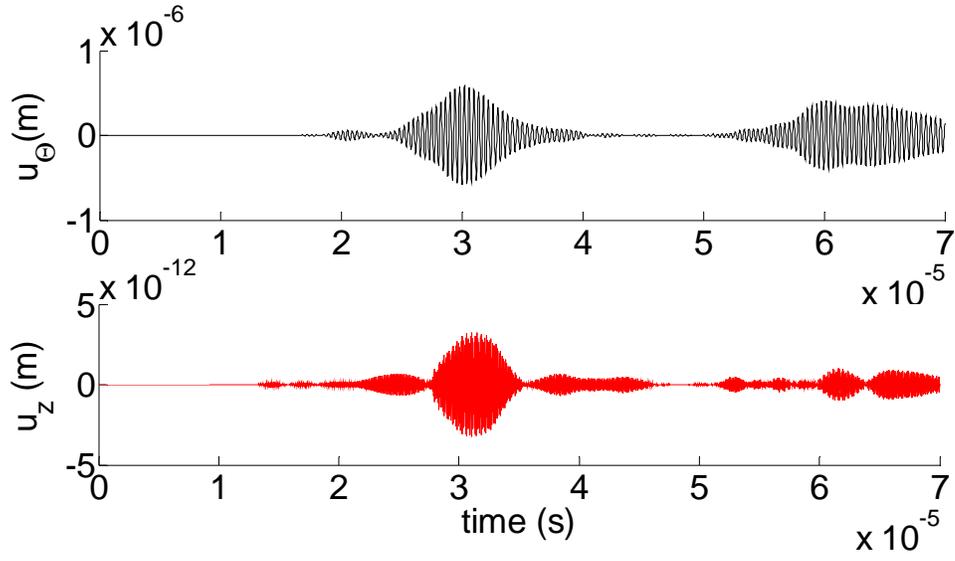

FIG. 4. (Color online) Time domain signals from $T(0,3)$ fundamental excitation at propagation distance of 57.5 *mm*. The black line is the displacement $u_\Theta$ component and the red line is the $u_z$ component.

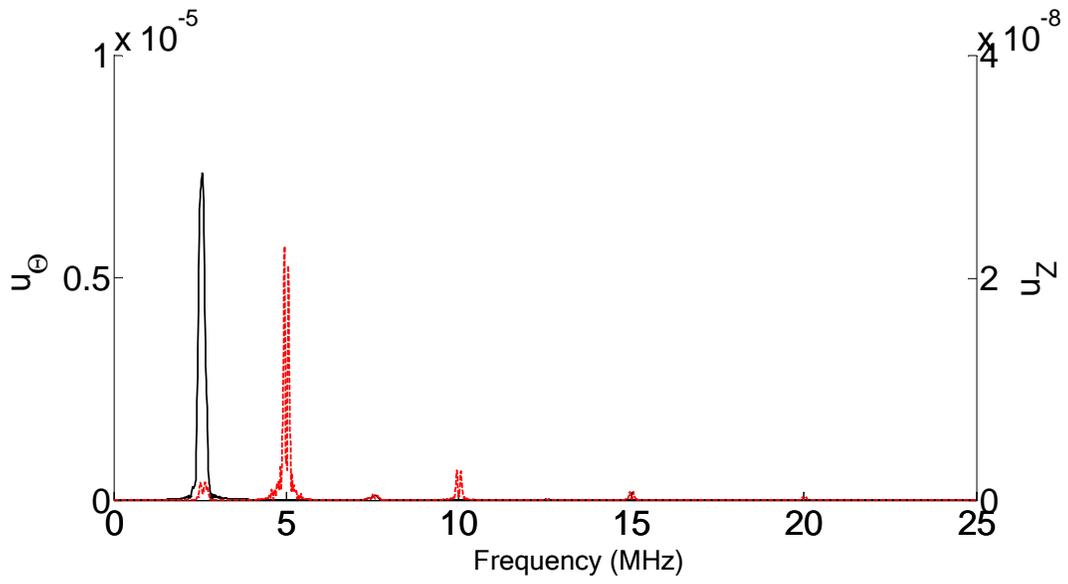

FIG. 5. (Color online) Frequency spectrums from $T(0,3)$ fundamental excitation: the black line shows the FFT result of the first receiving wave packet of the $u_\Theta$ displacement component, while the red dotted line indicates the frequency spectrum of the $u_z$ component.

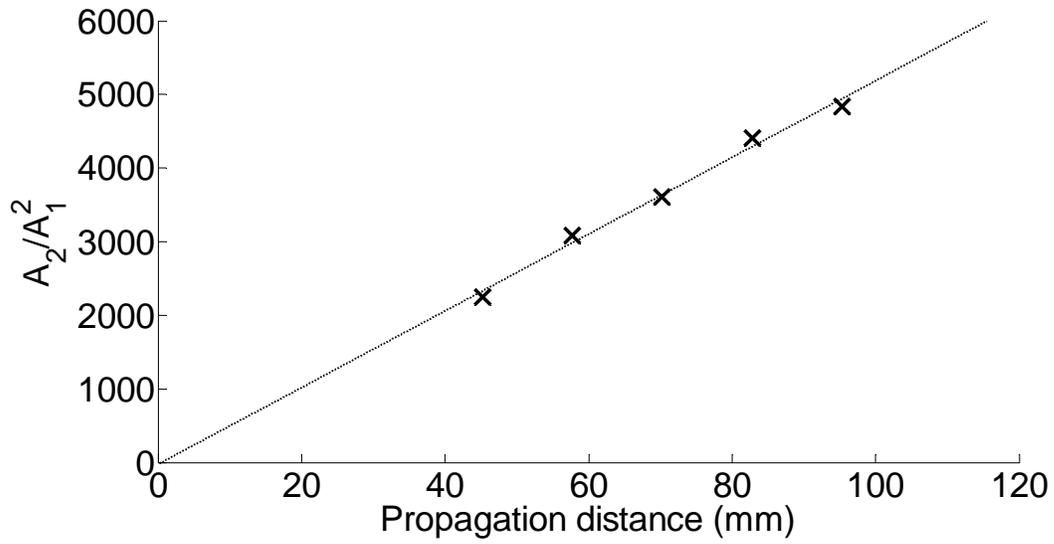

FIG. 6. The modal amplitude ratio $A_2/A_1^2$ for $T(0,3)$ mode excitation as a function of propagation distance. $A_1$ is the modal amplitude of $T(0,3)$ at fundamental frequency, while $A_2$ is the secondary $L(0,5)$ amplitude at double frequency. A best fit line of the data points is included.